\documentclass[doublecol]{epl2}

\newcommand{\be}{\begin{equation}} 
\newcommand{\ee}{\end{equation}}
\newcommand{\beqn}{\begin{eqnarray}} 
\newcommand{\eeqn}{\end{eqnarray}}

\title{The effect of asymmetric disorder on the diffusion in arbitrary networks}

\author{R\'obert Juh\'asz}
\institute{Institute for Solid
State Physics and Optics, Wigner Research Centre for Physics, H-1525 Budapest,
P.O. Box 49, Hungary}


\abstract{ 
Considering diffusion in the presence of asymmetric disorder,  
an exact relationship between the strength of weak disorder and 
the electric resistance of the corresponding resistor network is
revealed, which is valid in arbitrary networks. 
This implies that the dynamics are stable against weak asymmetric
disorder if the resistance exponent $\zeta$ of the network is negative. 
In the case of $\zeta>0$, 
numerical analyses of the mean first-passage time $\tau$ 
on various fractal lattices
show that the logarithmic scaling of $\tau$ with 
the distance $l$, $\ln\tau\sim l^{\psi}$, 
is a general rule, characterized by a new dynamical exponent $\psi$ 
of the underlying lattice. 
}

\begin{document}

\maketitle

\section{Introduction}
Disorder---an inevitable feature of nature---is known to induce  
striking slowing down phenomena in transport processes and
in the relaxation of 
systems with many degrees of freedom \cite{havlin,bouchaud,im}.
The source of complexity in such systems is twofold. 
First, the dynamics can be regarded as 
a random walk in the configuration space, which is, in general, 
 rather complicated. This problem also arises directly in 
the context of transport processes, which usually take place on
inhomogeneous structures in reality \cite{bouchaud}. 
Second, the system is frequently subject to an external 
source of disorder, such
as a random force-field, which can be modeled by quenched
(i.e. time-independent) random transition rates. 
The theoretical description of the dynamics of such systems 
is a challenging task 
since, in most cases, the complexity induced by disorder makes
the application of analytical techniques of clean systems extremely hard.  
Diffusion in the presence of the most general form of disorder, where 
transitions between pairs of states are non-symmetric, is non-trivial
even on regular $d$-dimensional lattices. 
In the simplest case of $d=1$, where a single-valued potential exists, 
the position $x(t)$ of the walker varies ultra-slowly with time, obeying    
\be 
\overline{\langle x^2(t)\rangle}\sim (\ln t)^{2/\psi},
\label{log}
\ee 
when the average force acting on the walker is zero \cite{sinai}.
Here, $\langle\cdot\rangle$ denotes an average over different stochastic 
histories, 
whereas the overbar denotes an average over the random transition rates. 
This is Sinai's diffusion law, where the value $\psi=1/2$ of the
barrier exponent is related to the Gaussian fluctuations of the potential
landscape.  
In higher dimensions, no potential can be defined, and the system
is a genuinely non-equilibrium process.
For increasing $d$, the effects of disorder are expected to be less pronounced
as there are more and more paths connecting the initial and final state.  
Weak disorder expansions \cite{dl} and perturbative renormalization group
analyses \cite{luck,fisher} agree in that, 
for $d\ge d_c=2$, the diffusion remains
normal, i.e. $\overline{\langle {\bf x}^2(t)\rangle}\sim t$, with logarithmic 
corrections at the critical dimension $d_c=2$, while for $d<2$ the disorder is
relevant and results in sub-diffusion.
Besides studying directly $x(t)$, the inverse question of how large the mean first-passage
time (MFPT) from a given initial state to a final one is frequently asked in
such systems \cite{redner}. In probing the dynamics, the finite-size scaling
of the MFPT is an alternative possibility as, in general, it obeys the same
dynamical relation between time and length scales as $x(t)$ does.  

In transition networks other than regular lattices neither a general
criterion for the relevance of asymmetric disorder 
nor the dynamics in case of relevance are known. 
Our aim in this letter is to study these questions in general.
We will prove an exact relationship between the strength of weak asymmetry in the transition
rates and the effective resistance of the corresponding resistor network, 
by which a relevance criterion can be formulated in terms of 
the sign of the resistance exponent $\zeta$.  
This will be then demonstrated by numerical calculations 
in various random and non-random fractal lattices. 
In the case of relevance ($\zeta>0$), the dynamics are found to be logarithmic 
given by Eq.(\ref{log}) with a barrier exponent $\psi$ that is characteristic 
of the underlying lattice.  

\section{The model and its renormalization}      
To formulate the above statements precisely, 
let us assume that the system has a finite number of states labeled 
by the integers
$i=1,2,\dots,N$ and consider a continuous-time random walk on them, 
where the transition
rates ($p_{ij},p_{ji}$) are from link to link independent random variables. 
The rates $p_{ij}$ and $p_{ji}$ are 
allowed to be different but their distributions are required to be
identical, so that  
the average local 'force' is zero, 
i.e. 
\be 
\overline{F_{ij}}\equiv\overline{\ln(p_{ij}/p_{ji})}=0
\label{symm}
\ee
on each link $(ij)$. 
For the sake of simplicity, we assume, furthermore, that 
every state is reachable from every other one. 
The central quantity in a fixed realization of the
transition network is the MFPT, $\tau_i^{(A)}$, which is
the expected value of the time needed to first reach any state in a fixed 
set $A$ of target sites when starting from state $i$. 
These quantities with different starting state $i$ obey the 
following backward master equation \cite{redner}: 
\be
\sum_ip_{ji}\tau_i^{(A)}-\sum_ip_{ji}\tau_j^{(A)}=-K_j,
\label{backward}
\ee
with $K_i=1$ for all $i$ and the boundary conditions $\tau_i^{(A)}=0$ for all
$i\in A$. 
As it has been recently pointed out in ref. \cite{monthus},
this form of the equation enables one to calculate $\tau_1^{(A)}$ 
by recursively eliminating all states other than $1$ and $A$ one after the
other in a way closely related to strong disorder
renormalization group (SDRG) methods \cite{im}. 
When eliminating state $k$, 
direct transitions between states that were linked with state $k$
are generated with the rates: 
\be 
\tilde p_{ij}^0 = p_{ik}p_{kj}/\sum_ip_{ki}.       \qquad (generation)  
\label{p_ruleA}
\ee 
If the link $(ij)$ has already existed before the elimination of $k$ than its
rates $p_{ij}$ are added to the newly generated ones, so 
that the final renormalized rates read as: 
\be 
\tilde p_{ij} = p_{ij} +\tilde p_{ij}^0.          \qquad (addition)  
\label{p_ruleB}
\ee 
The parameters $K_i$ at those states that were linked
with state $k$ are also renormalized 
as $\tilde K_i = K_i + p_{ik}K_k/\sum_ip_{ki}$. 
When all states except of $1$ and $A$ have been eliminated, the 
MFPT can be calculated by 
$\tau_1^{(A)}=\tilde K_1/\sum_{i\in A}\tilde p_{1i}$
\footnote{The denominator here has a direct probabilistic interpretation: 
$\sum_{i\in A}\tilde p_{1i}/\sum_{i}p_{1i}$ is the probability that 
the system, leaving state $1$, will reach state $A$ earlier than
state $1$.}.
In the simple case $d=1$, the generation rule in Eq. (\ref{p_ruleA}) 
is the only operation and its
product form results in a rapid decrease of effective rates under the
elimination (or renormalization) procedure that signals slow dynamics. 
In more complex networks, also the addition rule given in Eq. (\ref{p_ruleB}) 
is applied since alternative 
paths from state $i$ to $j$ which do not go through $k$ may exist. 
The more connected the network is the more frequently the addition occurs and the slower the effective rate decreases under renormalization. 

\section{Renormalization of the weakly asymmetric model}
In general, the recursions described above cannot be solved analytically. 
Nevertheless, when merely seeking an answer to whether disorder is
relevant, the analysis can be greatly simplified as follows. 
It is known in the case $d=1$ that any weak disorder
drives the system to the infinite-randomness fixed point of the SDRG 
\cite{im} that describes the logarithmic behavior in Eq. (\ref{log}).
Moreover, the behavior of the asymmetry in the effective transition rates 
alone reflects the logarithmic dynamics as 
$|\ln(\tilde p_{1N}/\tilde p_{N1})|\sim N^{1/2}$, 
and indicates the relevance of weak disorder. 
Therefore, we shall deal only with a weak, 
asymmetric perturbation of the symmetric model of
the form $p_{ij}/p_{ji}\equiv 1+\epsilon_{ij}$ with $\epsilon_{ij}$ 
infinitesimally small, and 
keep track of the renormalization of $p_{ij}$ and $\epsilon_{ij}$. 
The transformation rules then read in leading order in $\epsilon_{ij}$ as 
\beqn
\tilde\epsilon_{ij}^0 = \epsilon_{ik} + \epsilon_{kj},   \qquad (generation)  
\label{e_ruleA} \\
\tilde p_{ij}\tilde\epsilon_{ij} = p_{ij}\epsilon_{ij} + \tilde p_{ij}^0\tilde\epsilon_{ij}^0,   \qquad (addition)  
\label{e_ruleB}
\eeqn  
whereas the rates $p_{ij}$ still obey
Eqs. (\ref{p_ruleA}-\ref{p_ruleB}), however, they are
now symmetric, $p_{ij}=p_{ji}$ 
and transform identically to the reduction rules 
of a resistor network with resistances $r_{ij}\equiv 1/p_{ij}$ on links $(ij)$. 
Let us consider an ensemble of networks with fixed resistances $r_{ij}$ and random variables $\epsilon_{ij}$ on each link, for which 
Eq. (\ref{symm}) implies $\overline{\epsilon_{ij}}=0$, the overbar now
denoting the average over $\epsilon_{ij}$ on link $(ij)$.
In order to prove general statements we need  
to allow $\epsilon_{ij}$ to be non-identically distributed on different links
and to require that 
$\overline{\epsilon^2_{ij}}=\alpha r_{ij}$, where $\alpha$ is an
infinitesimally small global constant. 

\section{Relationship with the two-point resistance}
First, let us consider a special class of networks which
can be reduced to two fixed states, $a$ and $b$, 
by exclusively eliminating states with {\it two} links. 
An example is the hierarchical diamond lattice illustrated 
in Fig. \ref{fractal}. 
It is easy to see that, here, the generation and addition steps 
are equivalent to the well-known reduction of
resistors in series and in parallel, respectively, and follow the simple rules:
\beqn 
\tilde r=r_1+r_2, \quad \tilde\epsilon = \epsilon_{1} + \epsilon_{2}
\qquad \qquad (generation)  
\\
\tilde r^{-1}=r_1^{-1}+r_2^{-1}, 
\quad \tilde r^{-1}\tilde\epsilon = r_1^{-1}\epsilon_{1} + r_2^{-1}\epsilon_{2}
\quad (addition).
\eeqn
When any of the above operations is performed,
$\epsilon_{1}$ and  $\epsilon_{2}$ are always
independent, $\overline{\epsilon_1\epsilon_2}=0$, so one easily obtains that 
\be
\overline{\tilde\epsilon_{ij}^2}=\alpha\tilde r_{ij}
\label{relation}
\ee 
remains valid at any stage of the renormalization procedure, 
all the way to the last link connecting states $a$ and $b$: 
\be 
\overline{\tilde\epsilon_{ab}^2}=\alpha\tilde r_{ab}.
\label{last} 
\ee 

For arbitrary  networks, the elimination of states with more than two links 
cannot be avoided in general. When decimating a state with 
$n>2$ links, the relation 
Eq. (\ref{relation}) will be broken for the modified $n(n-1)/2$ links, 
moreover, their rates will become correlated with each other.  
Surprisingly, when the network is reduced to two (arbitrary) states, 
$a$ and $b$, the relationship 
in Eq. (\ref{last}) will still hold. 
In the following, it will be convenient to introduce the scaled asymmetry
parameters $\omega_{ij}\equiv p_{ij}\epsilon_{ij}$, 
which satisfy initially $\overline{\omega_{ij}^2}=\alpha p_{ij}$. 
It is sufficient to prove Eq. (\ref{last}) for complete networks (in which every state
is linked with every other one) and for an arbitrary set of 
rates $\{p_{ij}\}$,  
then it applies to any other network by  
setting $p_{ij}=\overline{\omega_{ij}^2}=0$ for appropriate links, 
which amounts to deleting that link. 
For $N=3$, the statement is true since this network 
belongs to the special class, 
while for $N>3$ we shall prove it by induction. 
Assume the Eq. (\ref{last}) holds for a fixed $N$. 
It follows from Eqs. (\ref{e_ruleA}-\ref{e_ruleB}) that the final 
asymmetry parameter is a linear combination of the initial ones: 
$\tilde \omega_{ab}^{(N)}(\{p_{ij}\})=\sum_{i<j}C_{ij}^{(N)}(\{p_{ij}\})\omega_{ij}$, where the
coefficients are functions of the set of rates and the summation goes over all
links. 
Using this, Eq. (\ref{last}) can be rewritten as 
$\sum_{i<j}[C_{ij}^{(N)}(\{p_{ij}\})]^2p_{ij}=\tilde p_{ab}^{(N)}(\{p_{ij}\})$. 
Now let us extend the network to a one state larger one and keep the 
notation of parameters on links of the $N$-state subgraph, while 
denoting the rates and the scaled asymmetry parameters on the links 
from the old state $i$ to the new one by $q_{i}$ and $\phi_i$,
$i=1,2,\dots,N$, respectively. 
If the new state is eliminated, then effective rates
$P_{ij}=q_iq_j/\sum_{i=1}^Nq_i$ and scaled asymmetry parameters 
$\Omega_{ij}=(q_j\phi_i-q_i\phi_j)/\sum_{i=1}^Nq_i$ are generated according to
Eqs. (\ref{p_ruleA}) and (\ref{e_ruleA}) at all old links, and are added to 
the old parameters as given in Eqs. (\ref{p_ruleB}) and (\ref{e_ruleB}).  
This yields for the parameters of the system with $N+1$ states:
\beqn
\tilde p_{ab}^{(N+1)}(\{p_{ij}\},\{q_{i}\})=
\textstyle\sum_{i<j}C_{ij}^2(p_{ij}+P_{ij}),
\label{pN1} \\
\tilde\omega_{ab}^{(N+1)}(\{p_{ij}\},\{q_i\})=\textstyle\sum_{i<j}C_{ij}(\omega_{ij}+\Omega_{ij}),
\label{omN1}
\eeqn
where we have used the shorthand notation 
$C_{ij}$ for $C_{ij}^{(N)}(\{p_{ij}+P_{ij}\})$. 
It is to be shown that the expected value of the 
square of Eq. (\ref{omN1}) is
equal to Eq. (\ref{pN1}) (multiplied by $\alpha$) for any $\{p_{ij}\},\{q_i\}$. 
Using that
$\{\omega_{ij}\}$ and $\{\Omega_{ij}\}$ are independent and 
$\overline{\omega_{ij}^2}=\alpha p_{ij}$, we obtain that 
this is equivalent to  
\be
\alpha\textstyle\sum_{i<j}C_{ij}^2P_{ij}=
\overline{(\textstyle \sum_{i<j}C_{ij}\Omega_{ij})^2}.
\label{lemma}
\ee
Expanding $\Omega_{ij}$ in terms of $\phi_i$ and using 
$\overline{\phi_i^2}=\alpha q_i$, 
the r.h.s. assumes the form
$\alpha\sum_{i=1}^NG_i^2q_i$ with the coefficients 
$G_i=\sum_{j\neq i}q_jC_{ij}/\sum_jq_j$, where $C_{ij}$ for $i>j$ is 
defined as $C_{ij}\equiv -C_{ji}$. 
Then Eq. (\ref{lemma}) can be rewritten after some algebra in the form 
\be
\sum_{i<j<k}q_iq_jq_k(C_{ij}+C_{jk}+C_{ki})^2=0,
\label{sumprod}
\ee
where the summation goes over all triangles of links. 
A sufficient condition for Eq. (\ref{sumprod}) to hold is that 
\be
C^{(N)}_{ij}(\{p_{ij}\})+C^{(N)}_{jk}(\{p_{ij}\})+C^{(N)}_{ki}(\{p_{ij}\})=0
\label{closed}
\ee
for all triangles $i<j<k$ (and, in fact, for all closed
paths) and for any set $\{p_{ij}\}$.  
Up to $N=4$ this can be justified by direct calculations. 
The key observation for the induction is that when extending the network by
adding a new state then the coefficients of the old links in the 
enlarged network are modified only by a shift in their argument as 
$C_{ij}^{(N)}(\{p_{ij}\})\to C_{ij}^{(N)}(\{p_{ij}+P_{ij}\})$, see 
Eq. (\ref{omN1}). Therefore the identity will hold in the enlarged network
for all closed paths consisting of old links. 
The same network with $N+1$ states can, however, be built from another
starting set of $N$ states by adding one extra state, as well, in which case 
the identity (\ref{closed}) can be extended to another set of links. 
It is easy to see that applying this
reasoning to all possible sets of $N$ states (note that $a$ and $b$ are always part of this set), the old links cover all possible
triangles of the extended network and the identity will be valid for $N+1$. 

So, we have proved that Eq. (\ref{last}) holds for arbitrary networks. 
This means that, although the two-point resistance depends 
on the particular network, 
the variance of the effective asymmetry parameter, which can be
interpreted as the strength of disorder, depends on the resistance 
{\it universally} at least for weak disorder. 
If the resistance is increasing(decreasing) with the system size then 
the effective asymmetry is getting stronger(weaker) on larger scales. 
So, the dynamical relation between time and length scale on lattices 
with a decreasing resistance 
is expected to be stable against weak disorder; otherwise,
it may be altered compared to that of the homogeneous system.   
Measuring the distance $l$ between $a$ and $b$ in terms of 
the linear size if the
network can be embedded in a $d$-dimensional lattice and in terms of the
length of shortest path otherwise, 
the resistance scales in many cases as $\tilde r_{ab}(l)\sim l^{\zeta}$, where
$\zeta$ is the resistance exponent. 
The strength of disorder then scales with $l$ in leading order as 
\be 
|\ln (\tilde p_{ab}/\tilde p_{ba})|=|\tilde\epsilon_{ab}| 
\sim \sqrt{\tilde r_{ab}}\sim l^{\zeta/2}. 
\ee
For $\zeta>0$ this is the same type of logarithmic behavior 
as known for $d=1$, with the barrier exponent $\psi^0=\zeta/2$. 
But unlike for $d=1$, 
this law will, in general, be modified for finite disorder 
by higher order correction terms which are
expected to be non-universal. Indeed, numerical results that will be presented 
in the rest of this work indicate, 
in the case of $\zeta>0$, a logarithmic scaling law but with $\psi\neq\psi^0$.    

The results obtained here are in accordance with previous ones on regular
lattices. 
For $d>2$, the two-point resistance tends to a constant as 
$\tilde r_{ab}(l)\sim l^{\zeta}+{\rm const}$, with
$\zeta=2-d<0$;  at $d=2$, where the diffusion is normal with a logarithmic
correction, it still increases but just barely (formally $\zeta=0$), 
while for $d=1$, $\zeta=1$.
So, in general, we expect disorder to be relevant(irrelevant) 
if $\zeta>0$($\zeta<0$) while, in the marginal case $\zeta=0$, 
it will induce at most corrections to the clean behavior.  
\section{Numerical analysis}
We have tested this criterion on fractal lattices, 
where, in the case of homogeneous rates, the law of diffusion is, in general, 
of the form $\langle{\bf x}^2(t)\rangle\sim t^{2/d_w}$, characterized by 
an anomalous diffusion exponent, $d_w>2$ \cite{havlin}. 
This problem has been thoroughly investigated but, apart from 
early Monte Carlo simulation studies \cite{pandey,mc}, 
not in the presence of asymmetric disorder.    
The Einstein relation between
diffusion and conduction implies that $\zeta=d_w-d_f$, where $d_f$ is the
fractal dimension. From this, one
can see that $\zeta>0$ for all fractals in $d=2$ and, in fact, 
$\zeta$ is positive for most fractals. 

\begin{figure}[h]
\onefigure[scale=0.5]{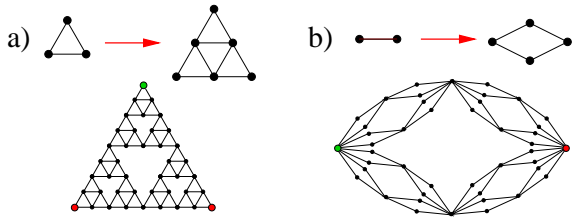}
\caption{
Iteration rule and the $4$th generation of the Sierpinski (a) and the
hierarchical diamond lattice (b). The starting(target) sites are denoted by
green(red) dots.    
}
\label{fractal} 
\end{figure}

\begin{figure}[h]
\onefigure[scale=0.6]{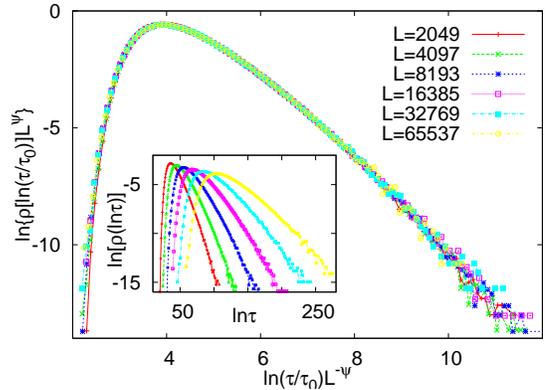}
\caption{Scaling plot of the distribution
  of the logarithm of the MFPT on Sierpinski triangles with different linear
  size $L$. The parameters are $\psi=0.296$ and $\ln\tau_0=-1.47$. The strength of disorder was $\lambda=0.1$. 
Inset: Unscaled distributions. 
}
\label{st}
\end{figure}

\begin{figure}[h]
\onefigure[scale=0.6]{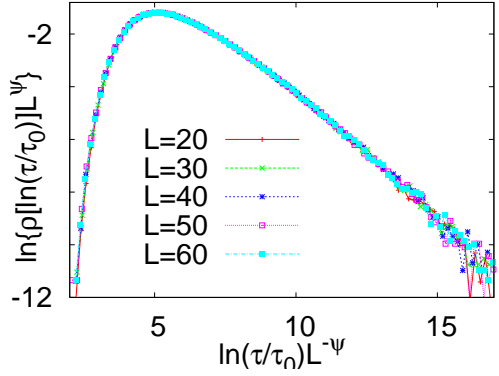}
\caption{The same plot for $d=3$ percolation
  clusters as in Fig. \ref{st} with $\psi=0.63$ and $\ln\tau_0=-5.7$.
}
\label{2d}
\end{figure}

We have numerically calculated the MFPT by recursively solving 
Eq. (\ref{backward}) on three different fractal lattices
with $\zeta>0$, namely, the Sierpinski triangle (ST) shown in
Fig. \ref{fractal} and bond percolation
clusters in $2$ and $3$ dimensions at the percolation threshold. 
In the former case, the MFPT from one of the tips of the
largest triangle to the other two tips was considered, whereas
in the latter case, percolating clusters that connect opposite
$d-1$-dimensional surfaces of a $d$-dimensional cube have been generated (with
periodic boundaries in the other directions), and the MFPT from
a given site on one surface to the opposite surface was considered.       
The rates on each link were $p_{ij}=1$, $p_{ji}=\lambda<1$ with a
random (equiprobable) orientation.  
The distributions of the logarithm of the MFPT determined  
in $10^6$ realizations for different
linear system sizes $L$ are found to broaden with $L$ in all three cases, 
see Figs. \ref{st}-\ref{2d} and to obey the scaling law 
\be
\rho[\ln(\tau/\tau_0),L]=L^{-\psi}\tilde\rho[\ln(\tau/\tau_0)L^{-\psi}],
\ee
where the barrier exponents $\psi$ are independent of the
strength of disorder $\lambda$, and are estimated for the 
ST and $d=2$ and $d=3$ percolation clusters to be, in order,  
$0.296(2)$, $0.46(2)$ and $0.63(1)$.  
\begin{figure}[h]
\onefigure[scale=0.6]{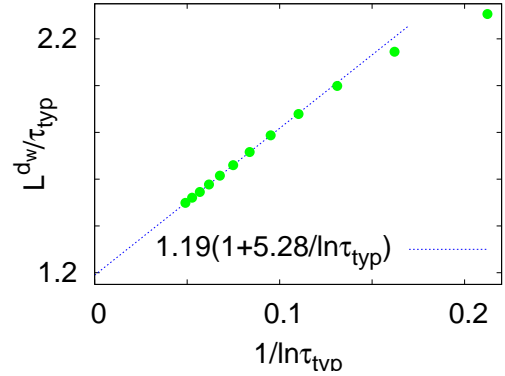}
\caption{
Finite-size behavior of the typical MFPT in the hierarchical diamond lattice 
with $\lambda=0.1$.
}
\label{hdl}
\end{figure}

We have the most accurate estimate for the ST, due to
an efficient renormalization, essentially by reversing the
construction procedure. Here, the barrier exponent $\psi=0.296(2)$ 
can be seen to be significantly different from 
$\psi_0=\zeta/2=\ln(5/3)/\ln 4\approx 0.368$. 

Concerning marginal structures ($\zeta=0$), we have considered the  
hierarchical diamond lattice (see Fig. \ref{fractal}), where $d_f=d_w=2$ and,
in fact, $\tilde r_{ab}=1$ in each generation.  
The distribution of $\tau/L^{d_w}$   
is found to converge to a limit
distribution for $L\to\infty$, but very slowly; 
the typical MFPT, $\tau_{\rm typ}\equiv \exp(\overline{\ln\tau})$, has 
a logarithmic correction of the same form as in $d=2$ regular lattices: 
$L^{d_w}/\tau_{\rm typ}\simeq D(1+a/\ln\tau_{\rm typ})$, see Fig. \ref{hdl}. 

\section{Discussion}
In summary, we have revealed a
close relationship between the effective strength of asymmetric disorder 
in the diffusion problem and the electric resistance. 
This  yields a simple relevance criterion 
in terms of the resistance exponent. 
If $\zeta<0$, the dynamics are stable against weak asymmetric disorder, 
while if $\zeta>0$,  they are unstable and, as numerical results
show, the logarithmic scaling is not a peculiarity of one dimension but 
it is the general rule, characterized by the exponent $\psi$ 
of the underlying structure that is independent of other 
known dynamical exponents. In the marginal case, $\zeta=0$, 
slow corrections to the clean system behavior are expected to be general.

As aforementioned, stochastic processes can be regarded as
random walks in their configuration spaces. 
In disordered stochastic systems with many degrees of freedom, 
logarithmically slow dynamical behavior indeed appears, such as in 
many-particle transport processes with zero-range or exclusion interaction 
\cite{jsi}, or in the contact process at criticality in one \cite{hiv} and
higher \cite{vfm} dimensions. 
The effect of disorder is, however, not restricted to appear in one single 
(critical) point of the parameter space of these systems (including the random walk in
one dimension) but the latter point is surrounded by an extended phase, called
Griffiths phase \cite{griffiths}, where the dynamical exponents are finite and vary with the
parameters of the model non-universally \cite{im}.   
In the random walk representation, the average force acting on the walker
becomes non-zero in some direction in this phase, therefore this is
out of the scope of the present treatment (cf. Eq. (\ref{symm})). 
Nevertheless, it is plausible to expect systems with a positive resistance 
exponent to have a Griffiths phase, 
when a biased force-field is applied.  
The non-universal behavior with a finite dynamical exponent is, however, 
suppressed if the lattice contains macroscopic 'dead-ends' (such as
in critical percolation clusters) from which the walker escapes after an
exponentially large trapping time \cite{havlin}, leading to
logarithmically slow dynamics also in the driven phase.          
The analyses of these phenomena is left for future research.

\acknowledgments
This work was supported by the J\'anos Bolyai Research Scholarship of the
Hungarian Academy of Sciences and by the National Research Fund
under grant no. K75324. 

\end{document}